\documentclass[12pt,preprint]{aastex}

\begin{document}

\title{A Comptonization Model for the Prompt Optical and Infrared Emission of GRB 041219A}

\author{Zheng Zheng\altaffilmark{1,2},Ye Lu\altaffilmark{1,3} and Yong-Heng Zhao\altaffilmark{1}}
\altaffiltext{1}{National Astronomical Observatories, Chinese
Academy of Sciences, 20A Datun Road, 100012 Beijing, China;
zz@bao.ac.cn} \altaffiltext{2}{Graduate University of Chinese
Academy of Sciences, 100080 Beijing, China}
\altaffiltext{3}{Department of Physics, The University of Hong
 Kong, Pokfulam Road, Hong Kong, China}

\begin{abstract}
Prompt optical emission from the $\gamma$-ray burst of GRB 041219A
has been reported by Vestrand et al. There was a fast rise of
optical emission simultaneous with the dominant $\gamma$-ray
pulse, and a tight correlation with the prompt $\gamma$-ray
emission has been displayed. These indicate that the prompt
optical emission and $\gamma$-ray emission would naturally have a
common origin. We propose that this optical component can be
modeled by considering the Comptonization of $\gamma$-ray photons
by an electron cloud. As a result of this mechanism, the arrival
time of the optical photons is delayed compared with that of the
$\gamma$-rays. We restrict that the lag time to be shorter than
$10$ s, within which the prompt optical emission is considered to
vary simultaneously with the prompt $\gamma$-ray emission. Taking
the observations of GRB 041219A into account, we derive the number
density of the surrounding electron cloud required for
Comptonization. The redshift of GRB 041219A is predicted to be
$z\lesssim 0.073$ as well.
\end{abstract}


\section{Introduction}
GRB 041219A was detected by both the IBIS detector on the {\rm
International Gamma-Ray Astrophysics Laboratory} satellite
\citep{got04} and the Swift Burst Alert Telescope (BAT)
\citep{bar04}. It was an unusually bright and long burst. The
fluence of $1.55\times 10^{-4}\,ergs\,cm^{-2}$ measured by BAT in
the 15-350\,keV band  would put it among the top few per cent of
the 1637 $\gamma$-ray burst (GRB) events listed in the
comprehensive fourth BATSE (Burst and Transient Source Experiment)
catalog \citep{pac99}. The duration of the prompt $\gamma$-ray
emission was approximately $520\,s$, making it one of the longest
bursts ever detected \citep{ves05}.

So far, prompt optical and infrared emission that occurs when the
main burst is still in progress has been detected from a few GRBs.
These include GRB 990123 \citep{ake99}, GRB 041219A \citep{ves05,
bla05}, GRB 050401 \citep{ryk05}, GRB 060111B (Klotz et al. 2006),
and GRB 050904 \citep{Boe06, Wei06}. The data obtained by RAPTOR
(Rapid Telescopes for Optical Response) shows that the prompt
optical light curve of GRB 041219A could be well fitted with a
constant prompt optical-to-$\gamma$-ray flux ratio
$F_{opt}/F_{\gamma}=1.2\times 10^{-5}$ \citep{ves05}. This
strongly suggests that a direct correlation of both the
time-varying spectral shape and the flux magnitude exists between
the prompt optical emission and the $\gamma$-ray emission.

Prompt long-wavelength radiation accompanying prompt $\gamma$-ray
emission has been widely discussed by many authors in the
pre-afterglow era \citep{kat94, sch94, wei97, tav96,zha05} and in
the afterglow era \citep{Sai99, mes99, wu06, van00, fan04, fw04,
Bel05}. Recently, simultaneous variation of the optical-infrared
emission with the prompt $\gamma$-rays from GRB 041219A has been
discussed by \cite{fan05} using a neutron-rich internal shock
model. Alternatively, here we argue that Compton attenuation of
the $\gamma$-ray photons with a power-law spectrum by intervening
electron clouds could give birth to the prompt optical emission.
The $\gamma$-ray photons are assumed to be produced by the central
engine (e.g., standard fireball model), while an electron cloud
with the extremely high number density required could likely be
ejected by the progenitor of the GRB, for example, as ejecta from
the associated supernova in an earlier phase of the explosion
\citep{mac99}. When the incident $\gamma$-ray photons travel
through the electron cloud, some of them will be reprocessed into
optical photons through the Compton attenuation mechanism. The
model is addressed in detail in \S\, 2, and we give the discussion
and conclusions in \S\, 3.

\section{Model description for the prompt optical emission of GRB 041219A}

The distribution of the electron cloud is presumed to be
inhomogeneous. This situation is possible if the electron cloud
was created by a supernova that is associated with the GRB
(further details are given below). When $\gamma$-ray photons
escape from the central engine (e.g., a fireball) and encounter
the surrounding electron cloud, energy can be exchanged between
photons and electrons by Compton scattering. This process is often
referred to as {\rm $Comptonization$}. It is of vital importance
to know the environment of the Comptonization. If the scattering
happens in a low electron density region, the resulting emission
will be a GRB. Otherwise, saturated Comptonization must occur, and
most of the energy of the incident $\gamma$-ray photons will be
transferred into the electrons through repeated Compton
scatterings. This makes it possible for the $\gamma$-ray photons
to eventually be attenuated to the prompt optical-infrared
emission (see Fig. 1).

To demonstrate how energy can be interchanged between photons and
electrons by Compton scattering in detail is very complicated
except in some simplified limiting cases. We thus introduce the
Monte Carlo method to treat this problem. The simple picture can
be described as follows: We initiate the Comptonization process
with a photon of energy $\epsilon_0$, located at the coordinate
origin. When the photon travels away from its initial location by
a length equal to the free path $\lambda$, one scattering will
occur, and the scattering angle is $\alpha$. Assuming the
coordinate system is built along our sight line, then the initial
state of the incident photon is labeled as $P_0=(\epsilon_0,
\lambda_0, 0)$. We assume that the first scattering, with angle
$\alpha_1$, change the initial energy $\epsilon_0$ to $\epsilon_1$
over a length $\lambda_1$. After the first scattering, the state
of the photon can be labeled as $P_1=(\epsilon_1, \lambda_1,
\alpha_1)$. We label the next scattering state as
$P_2=(\epsilon_2, \lambda_2, \alpha_2)$, and so on. We continue
the Comptonization process until the incident photon escapes from
the cloud, and the final scattering state is
$P_m=(\epsilon_m,\lambda_m, \alpha_m)$. Thus the whole history of
the incident photon before it escaped from the electron cloud can
be described by
\begin{eqnarray}
P_0\rightarrow P_1\rightarrow P_2...\rightarrow P_j...\rightarrow
P_m\,\nonumber
\end{eqnarray}

Assuming that the scattering region has an electron density of
$n_e$ and a size of $L$, the free path of a photon with energy
$\epsilon_j$ is
\begin{equation}\lambda_j=\sigma^{-1}_j n_e^{-1} ln\xi\,\,\,,
\end{equation}
where $\xi$ is a random variable that is uniformly distributed in
(0,1), $\sigma_j$ is the cross-section determined by integrating
the differential cross-section $d\sigma_j$ over scattering angles
$\alpha_j$. We use the Klein-Nishina formula to calculate
$d\sigma_j$:
\begin{equation}
d\sigma_j=\pi
r_{0}^2\frac{\epsilon_{j+1}^2}{\epsilon_j^2}(\frac{\epsilon_{j+1}}{\epsilon_j}+\frac{\epsilon_j}{\epsilon_{j+1}}-\sin
^2\alpha_j)\sin \alpha_j d\alpha_j
\end{equation}
\citep{ryb79}, where $r_{0}$ is the classical electron radius and
$\epsilon_j$ and $\epsilon_{j+1}$ are the energy of any incident
photon and the next scattering photon, respectively. The relative
energy change of the photon satisfies
\begin{equation}
\frac{\epsilon_{j+1}}{\epsilon_j}=\left[1+\frac{\epsilon_j}{m_{e}c^{2}}(1-\cos\alpha_j)\right]^{-1}\,\,,
\end{equation}
where $m_{e}$ is the mass of the electron. Consequently, the total
optical distance traveled by one single photon for the repeated
scatterings is the sum of the free paths, that is,
\begin{eqnarray}
S=\sum\lambda_j\,\,\,.
\end{eqnarray}
Because $\lambda_j \ll L$, the photon does not travel through the
electron cloud along a straight line. But, the sum of the
projection of each scattering length $\lambda_j$ onto the sight
line equals the size of the scattering region, that is,
\begin{eqnarray}
L=\sum\lambda_j\cos\theta_{j-1}\,\,\,,
\end{eqnarray}
where $\theta_j$ is the separation angle between the scattering
direction and the sight line. It is determined by
\begin{eqnarray}
\cos\theta_{j}=\cos\theta_{j-1}\cos\alpha_j+\sin\theta_{j-1}\sin\alpha_j\cos\phi\,\,,
\end{eqnarray}
where $\phi$ is a random variable that is uniformly distributed in
($0,2\pi$).

To carry out the simulations, we need to fix the election cloud's
length $L$ and the electron number density $n_e$. Two scenarios
for the scattering region of the electron clouds are considered:
One is that the total scattering area of the electron cloud is
assumed to be covered by a high and a low electron density region,
referred to as the HL case. The geometry of the electron clouds
for this picture is plotted in Figure 1. Another is that the
scattering electron cloud could have only a high density filled in
with some vacuum gaps, referred to as the HG case.

We first estimate the values of $L$ and $n_e$ in the HL case. We
simulate $10^3$ $\gamma$-ray photons with incident energy of
$200$\,keV passing through the high electron density region, and
require that the final radiation due to Comptonization is in the
prompt optical band with energy of $2\,eV$. Combining equations
(1)-(6), we obtain the statistical quantities $Ln_{e}$ and
$S_{h}n_{e}$ from the simulations:
\begin{eqnarray}
Ln_{e}=4.37\times 10^{26}\,cm^{-2},\,\,\, S_hn_{e}=1.227\times
10^{29}\,cm^{-2}\,\,\,,\nonumber
\end{eqnarray}
where $S_{h}$ is the total optical distance of a photon that
passes through the relevant high-density region of the electron
cloud. Correspondingly, we use $S_{l}$ to denote the total optical
distance of a photon that passes through the low-density region,
and the relation $S_{h}\gg S_{l}$ can be satisfied by taking into
account that the value of $n_{e}$ in the high-density region is
far more than that of the low-density region.

It is encouraging that the emergence of a weaker component after
the end of the prompt $\gamma$-ray emission, detected in the
PAIRITEL near-infrared observation of GRB 041219A, is interpreted
as delayed reverse shock emission \citep{bla05}. As a result of
the $Componization$ mechanism, the optical emission delays that of
the $\gamma$-rays. This is favored by the work of \citet{Tan06},
who independently arrived at the conclusion that the prompt
optical emission from GRB 041219A is consistent with the a delay
of several seconds with respect to the $\gamma$-ray emission in
their rest frames, if the redshift of GRB 041219A is taken to be
0.1 \citep{Bar05}. We restrict the prompt optical emission of GRB
041219A to lag by less than $\delta t\sim 10$\,s with respect to
the $\gamma$-ray emission, within which it is roughly regarded as
varying with the prompt $\gamma$-ray emission. If the evolution of
the spectrum of GRB 041219A is primarily determined by Compton
scatterings, then we have $ S_{h}-S_l=c\delta t \leq 3\times
10^{11}\,cm,$ where $c$ is the velocity of light.

With $S_h\gg S_l$, we can immediately derive the number density of
electrons required for the incident $\gamma$-ray photons to be
degraded to optical photons through $Comptonization$, $ n_e \geq
4.07\times 10^{17}\,cm^{-3}. $ The corresponding size of the
scattering region is also estimated as $L\leq 1.07\times 10^{9}\,
cm$.

The real situation is that the observed prompt optical emission
from GRB 041219A would likely only be composed of a saturated
Comptonization component from the total incident $\gamma$-ray
photons. Note that the spectrum of GRBs is nonthermal, and there
is a long high-energy tail extending up to GeV levels, with the
energy flux peaking at a few hundred keV in many bursts
\citep{pir05}. For  simplicity, we assume that the incident energy
distribution of $\gamma$-ray photons follows a power law with
 index $p=-1$ and use $10^5$ $\gamma$-ray photons with energy
varying from $100\,MeV$ to $10\,keV$ to test our model. Fixing the
size of the scattering region at $L\approx5.6\times10^8 cm$, we
find that if the number density of the electron cloud is
$n_e\approx 7.2\times10^{17}cm^{-3}$, these incident $\gamma$-ray
photons mostly become UV, optical, and infrared photons after the
Comptonization process. On the contrary, if they pass through an
electron cloud with $n_{e}\approx2.6\times10^{16}\,cm^{-3}$, the
emergent photons are mostly in the $\gamma$-ray band with energies
of $10-10^3\,keV$, and the distribution of these emitted
$\gamma$-rays obeys the same power law as the initial incident
photons. For these two conditions, we have simulated the process
by setting the total number of incident photons at $10^5$. The
total energy carried by these photons is assumed to be
$\sim8.5\times10^{9}\,eV$. After Comptonization, the emerging
photons traveling along the line of sight are markedly different
in the former case and in the latter case, respectively. In the
former case, there are photons are about 697 escaped photons, of
which 34 are emitted in the IR/optical band (1.7-2.2\,eV), with an
energy of $\sim67\,eV$. In the latter case, about 16,744 escape,
of which 536 are in the $\gamma$-ray band (15-350\,keV), with a
total energy of $\sim3.8\times10^7eV$. The simulation for the
numbers of radiated photons per energy interval is plotted in
Figure 2.

To fit the energy flux ratio between the optical emission and the
prompt $\gamma$-ray emission, we introduce a covering factor $f$,
which is defined as $f=A_{h}/A_{tot},$ where $A_{h}$ is the area
of the high-density clouds covered and $A_{tot}$ stands for the
total area of the electron clouds. This could be the sum of the
high- and low-density regions, respectively (see Fig. 1).

Given the covering factor $f$, from the simulations one can
immediately derive the flux ratio of the optical to $\gamma$-rays,
$R=67f/[3.8\times10^7(1-f)]$. Note that the observation of GRB
041219A \citep{ves05} shows that the flux ratio between the prompt
optical emission (1.7-2.2\,eV) and the prompt $\gamma$-ray
emission (15-350\,keV) is $F_{opt}/F_{\gamma}=1.2\times 10^{-5}$.
By fitting the observation with $R=F_{opt}/F_{\gamma}$, we derive
 $f \sim 87\%$. The simulation also demonstrates the fact that the total
emergent $\gamma$-ray (15-350\,keV) photon energy is about
$f_\gamma^{out}\sim 0.4\%$ of the total incident photon energy
(see Fig. 2).

By considering the average effect of  vacuum gaps that fill in the
high-density region and roughly replace the low-density region of
the HL case, the calculations and simulation show that a covering
factor of more than $99\%$ is required to fit the observational
data in the HG case. Compared with the HL case, this covering
factor is extremely high. We do not think that so few vacuum gaps
would be feasible, since nearly all the area is taken up by the
high-density region. Therefore, the results for the HG case are
not included, and we do not concern ourselves with this scenario
any further.

It is known that the fluence of GRB 041219A in the 15-350\,keV
energy band measured by BAT \citep{bar04} is $F_\gamma\sim
1.55\times 10^{-4}\,ergs\,cm^{-2}$. With the isotropic energy of
GRB 041219A, one can estimate its red-shift from the formulae
\begin{eqnarray}
&& f^{out}_\gamma E_{iso}(1-f)=4\pi d^2F_\gamma\,\,\,,\nonumber\\
&& H_{0}d=(1+z)\int_{0}^z
[(1+x)^2(1+x\Omega_{M})-x(2+x)\Omega_{\Lambda}]^{-1/2}dx,
\end{eqnarray}
where $d$ is the luminosity distance between the GRB and us
\citep{Carr92}, $x$ is an integral variable for the redshift, and
$\Omega_M$, $\Omega_\Lambda$, and $H_0$ are the cosmological
parameters and Hubble constant, respectively. If the isotropic
energy of GRB 041219A is less than that of GRB 990123,
$E_{iso}\sim 3.4\times 10^{54}\,ergs$ \citep{kul99}, then by
combining equation(7), $H_0=75\,km\,s^{-1}Mpc^{-1}$,
$\Omega_{M}=0.3$, $\Omega_{\Lambda}=0.7$, and $f=87\%$, we obtain
an upper limit for the redshift of $z \lesssim 0.073$, which is
roughly consistent with the result of \citet{Bar05}.

\section{Discussion and Conclusion}
We propose that the prompt optical and infrared emission from GRB
041219A can be modeled by the saturated Comptonization of its
$\gamma$-ray emission. The current evidence shows that the optical
emission of GRB 041219A is rather distinct from that of other
GRBs, suggesting a possible environmental influence \citep{ves05}.
We argue that this special event could have been surrounded by a
very high density electron cloud. If the sub-$10\,s$ lag time
prompt optical emission can be considered as varying
simultaneously with the prompt $\gamma$-ray emission, by fitting
the observations we find that the required electron number density
to degrade the $\gamma$-rays into the optical band is $\sim
7\times 10^{17}\,cm^{-3}$, and this dense scattering region is
expected to cover $\sim 87\%$ of the total area. Although the
required value of the electron number density is much higher than
that of ordinary interstellar matter, it is lower than the
electron number density of the Sun's outer convection layer
\citep{bah88}.  We argue that the electron cloud could be a result
of, for example, either the ejecta of the outer layer of the GRB
progenitor ejected in an earlier phase of the explosion
\citep{mac99} or the capture of a star by the GRB progenitor
\citep{wan01}. If the scattering electron cloud is produced by the
former mechanism, its total mass can be estimated as $M\sim 4\pi
R_{\ast}^2Ln_{e}m_{p}\approx 2 \times10^{-6}\,M_{\odot}$, where
$m_{p}$ is the proton mass and $R_*$ is the radius of the GRB
progenitor star, with $R_*\sim10R_{\odot}$ adopted here. Suppose
that the ejected outer layer of the progenitor has a typical
length of $10^5\,cm$ and density of $10^{-3}\,g\,cm^{-3}$; when it
expands into a size of $10^8\,cm$, it could create an electron
cloud as required by the model. As a result, we predict that an
upper limit on the redshift of GRB 041219A of $z\sim 0.073$ if the
isotropic energy of this GRB is less than that of GRB 990123.

Further more, many photons radiated in the UV/soft X-ray band
could be inferred. Nevertheless, the XRT and UVOT instruments on
board \emph{Swift} did not autonomously slew to the burst, since
automated slewing is not yet enabled \citep{fen04}; it is
interesting to note that the \emph{Rossi X-Ray Timing Explorer}
All Sky Monitor  obtained some soft X-ray data for the initial
120\,s of GRB 041219A \citep{mcb06}. The results presented here
highlight the need for continued broadband observations of
$\gamma$-ray burst and the afterglow.

\noindent{\bf Acknowledgement:} We thank the anonymous referee for
valuable comments and suggestions that lead to an overall
improvement of this study. We are thankful to Y.-F. Huang, S.-N.
Zhang and K.-S. Cheng for both helpful comments and useful
discussions. Thanks also to H.-X. Yin and W. Qiao for useful
discussions. This research was supported by the National Natural
Science Foundation of China (grants 10573021, 10273011, and
10433010), and by the Special Funds for Major State Basic Research
Projects.


\clearpage
\begin{figure}
\includegraphics[scale=0.3]{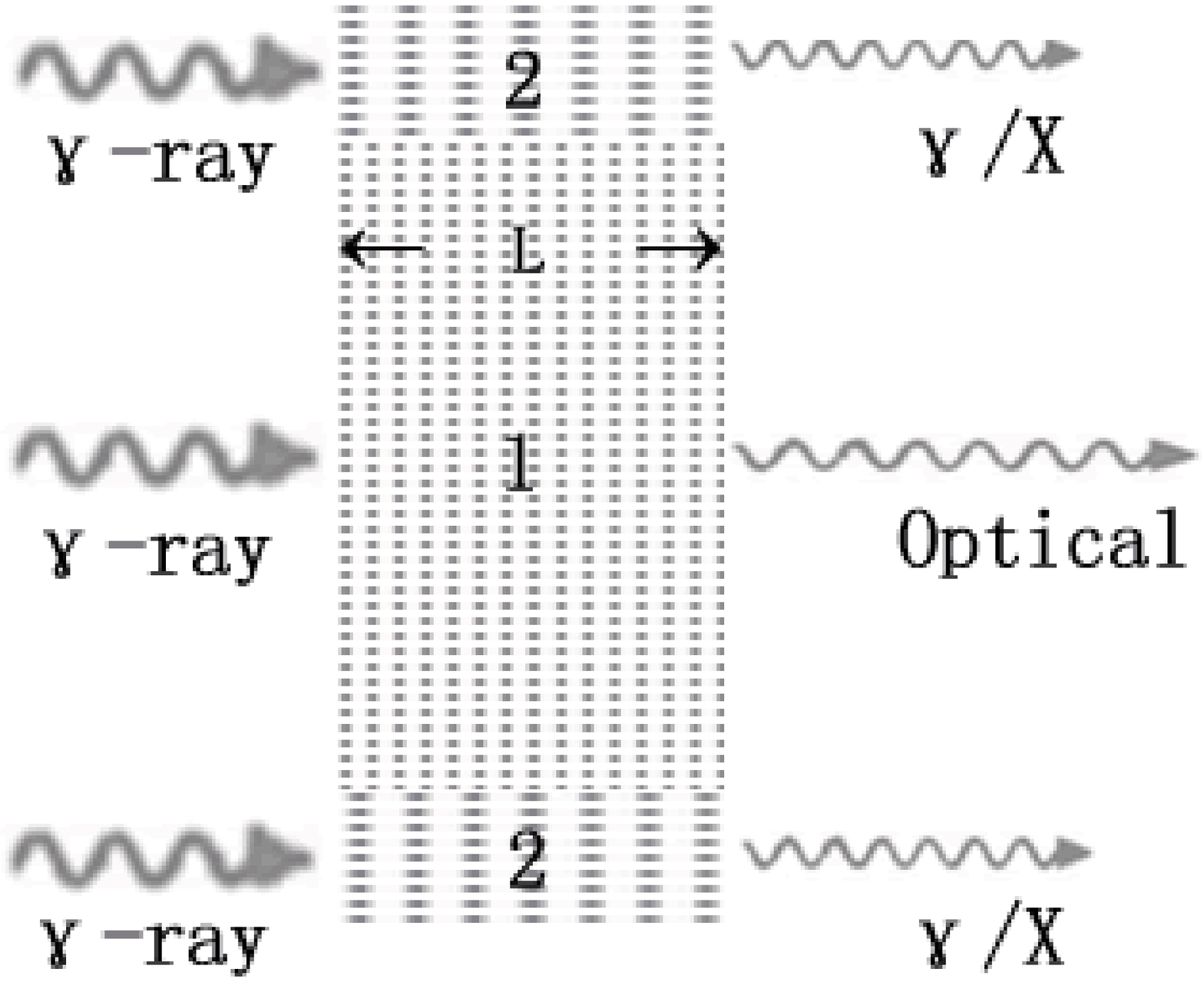} \caption{Geometry of the electron clouds
surrounding the central $\gamma$-ray source. The total area of the
electron clouds is separated into two kinds of regions with
different electron densities: Region 1 corresponds to the area
covered by high density, and region 2 stands for the area covered
by low density. The length of the clouds is denoted by L.}
\end{figure}

\clearpage
\begin{figure}
\plottwo{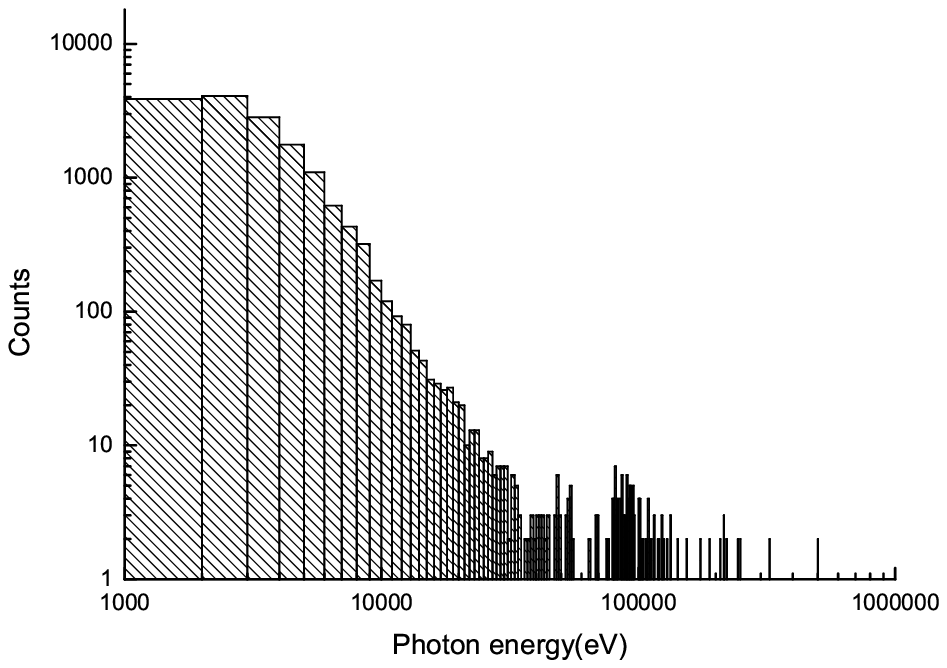}{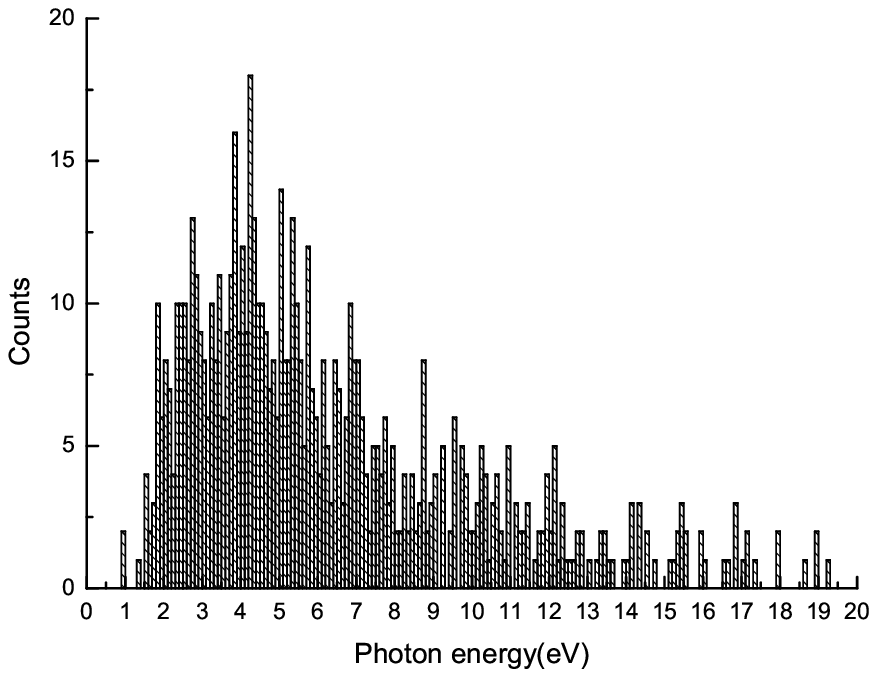} \label{sfrdez} \caption{Spectrum of
radiated photons that pass through an electron cloud with number
density of $2.6\times10^{16}\,cm^{-3}$ (\emph{left}) and
$7.2\times10^{17}\,cm^{-3}$ (\emph{right}). The total number of
incident photons  is taken to be $10^5$, and the numbers of
emitted photons are 16,744 (of which 536 are in the $\gamma$-ray
band) in the left panel, and 697 (of which 34 are in the optical
band) in the right panel. The counts of the radiated photons are
estimated with a proportion of $\epsilon^{-2}$.}
\end{figure}


\begin{thebibliography}

\bibitem[Akerlof et al.(1999)]{ake99}Akerlof, C., et al., 1999,
\nat, 398, 400
\bibitem[Barkov \& Bisnovatyi-Kogan(2005)]{Bar05} Barkov, M. V., \& Bisnovatyi-Kogan, G. S. 2005, Astrophysics,
48, 369
\bibitem[Bahcall \& Ulrich(1988)]{bah88}Bahcall, J.N., Ulrich, R.K., 1988, Rev. Modern
Phys., 60, 297
\bibitem[Barthelmy et al.(2004)]{bar04}Barthelmy, S., et al., 2004, GCN Circ.
2874, 

 \bibitem[Beloborodov(2005)]{Bel05}Beloborodov, A.M., 2005, \apj, 618,
 L13

 \bibitem[Boer et al.(2006)]{Boe06}Bo\"{e}r, M., Atteia, J. L., Damerdji, Y., Gendre, B.,
 Klotz, A., \& Stratta, G., 2006, \apj, 638,
L71

\bibitem[Blake et al.(2005)]{bla05}Blake, C.H., {\it et al.} 2005, \nat, 435, 181

\bibitem[Carroll et al.(1992)]{Carr92}Carroll, S. M., Press, W. H., \& Turner, E. L., 1992, ARA\&A, 30, 499-542

\bibitem[Fan \& Wei(2004a)]{fan04}Fan, Y.Z., \& Wei, D.M., 2004,
\mnras, 351, 292
\bibitem[Fan \& Wei(2004b)]{fw04}Fan, Y.Z., \& Wei, D.M., 2004,
\apj, 615, L69

\bibitem[Fan et al.(2005)]{fan05}Fan, Y.Z., Zhang, B. \& Wei, D.M., 2005,
\apj, 628, L25
\bibitem[Fenimore et al.(2004)]{fen04}Fenimore, E., et al, 2004,
GCN Circ. 2906, 
\bibitem[Gotz et al.(2004)]{got04}Gotz,D., Mereghetti,S., Shaw, S., Beck, M., \& Borkowski, J.
2004, GCN Circ.2866, 
\bibitem[Katz(1994)]{kat94}Katz, J.I., 1994, \apj, 432, L107
\bibitem[Klotz et al.(2006)]{Klo06} Klotz, A., et al., 2006, astro-ph/0604061
\bibitem[Kulkarni et al.(1999)]{kul99} Kulkarni, S.R., et al., 1999, Nature, 398, 389
\bibitem[MacFadyen \& Woosley(1999)]{mac99}MacFadyen,A.I. \& Woosley,S.E., 1999, \apj, 524,
262

\bibitem[McBreen et al.(2006)]{mcb06}McBreen, S., et al., 2006,
astro-ph/0604455, accepted for publication in A\&A
\bibitem[M\'{e}sz\'{a}ros \& Rees(1999)]{mes99}M\'{e}sz\'{a}ros, P., \&
Rees, M.J., 1999, \mnras, 306, L39

\bibitem[Paciesas et al.(1999)]{pac99}Paciesas, W., et al., 1999, ApJS, 122, 465
\bibitem[Piran(2005)]{pir05}Piran, T., 2005, Rev. of Mod. Phys., 76, 1143

\bibitem[Rybicki \& Lightman(1979)]{ryb79}Rybicki, G. B. \& Lightman A.P., 1979, "Radiative Progresses In
Astrophysics", John Wiley \& Sons Inc

\bibitem[Rykoff et al.(2005)]{ryk05}Rykoff, E.S., et al., 2005,
\apj, 631, L121
\bibitem[Schaefer et al.(1994)]{sch94}Schaefer, B.E., 1994, \apj,
422, 71

\bibitem[Sari \& Piran(1999)]{Sai99}Sari, R., \& Piran, T., 1999,
\apj, 517, L109

\bibitem[Tang \& Zhang (2006)]{Tan06}Tang, S.M., \& Zhang, S.N., 2006, submitted to
A\&A
\bibitem[Tavani(1996)]{tav96}Tavani, M., 1996, \apj, 466, 768
\bibitem[Vestrand et al.(2005)]{ves05}Vestrand, W.T., et al. 2005, \nat, 435, 178
\bibitem[van Paradijs et al.(2000)]{van00}van Paradijs, J., Kouveliotou, C., Wijers R. A. M.
J., 2000, ARA\&A, 38, 379
\bibitem[Wang \& Zhao(2001)]{wan01}Wei W. \& Zhao, Y.H., 2001,
ChJAA, 6, 487

\bibitem[Wei et al.(2006)]{Wei06}Wei, D. M., Yan, T., \& Fan, Y.
Z., 2006, \apj, 636, L69

\bibitem[Wei \& Cheng(1997)]{wei97}Wei, D.M., \& Cheng, K.S.,
1997, \mnras, 290, 107

\bibitem[Wu et al.(2006)]{wu06} Wu, X.F., Dai, Z.G., Wang, X.Y., Huang, Y.F.,
Feng, L.L., Lu, T., 2006, ApJ, submitted (astro-ph/0512555)

\bibitem[Zhang(2005)]{zha05}Zhang, B., 2005, to appear in Proc. of
"Astrophysics Sources of High Energy Particles and Radiation",
eds. T. Bulik, G. Madejski and B. Rudak, Torun, Poland, 20-24,June
(astro-ph/0509571)

\end{thebibliography}
\end{document}